\documentclass[aps,prl,twocolumn,floatfix]{revtex4-2}
\usepackage{amsmath,amssymb,bm}
\usepackage{booktabs,array}
\usepackage[colorlinks=true,linkcolor=blue,citecolor=blue,urlcolor=blue]{hyperref}

\begin{document}
\title{Neutrino Spin--Flavor Dynamics and Quantum Rabi Model: Spectral Correspondence and QRM Benchmarks}
\author{S. Zarrinkamar}
\affiliation{Department of Basic Sciences, Ga. C., Islamic Azad University, Garmsar, Iran}
\email{saber.zarrinkamar@iau.ac.ir, zarrinkamar.s@gmail.com}
\begin{abstract}
We formulate a two-state neutrino spin--flavor system coupled through a transition magnetic moment to a single quantized electromagnetic mode and establish its operator-level correspondence with the quantum Rabi model (QRM). For the sector $\{\lvert\nu_{iL}\rangle,\lvert\nu_{jR}\rangle\}$, the diagonal splitting is $\Delta_s=(m_i^2-m_j^2)/(2E)+V_{iL}-V_{jR}$, while quantization of a transverse electromagnetic mode gives a coupling $g_\nu=\mu_{ij}B_{\rm zpf}$. The effective Hamiltonian consequently has the QRM form $H_\nu=\Delta_s\sigma_z/2+\omega_c a^\dagger a+g_\nu\sigma_x(a+a^\dagger)$. We then formulate the correspondence at the level of the exact QRM spectrum, focusing on its exceptional (Juddian) states, and transfer the QRM constraint polynomials directly to neutrino variables. This yields explicit algebraic conditions on the neutrino mass splitting, energy, transition magnetic moment, zero-point magnetic field, and mode frequency. We also identify QRM phenomena that would become conditional predictions of the quantized-field neutrino model---including counter-rotating corrections, parity structure, spin--field entanglement, and occupation-dependent dynamics. The trapped-ion QRM experiments provide benchmarks for the operator family, while none of these QRM phenomena is presented as an already observed neutrino effect.
\end{abstract}
\maketitle
\section{Model and correspondence}
The quantum Rabi model is the canonical Hamiltonian for a two-level system coupled to a quantized bosonic mode \cite{Rabi1937},
\begin{equation}
 H_{\rm QRM}=\frac{\omega_0}{2}\sigma_z+\omega_c a^\dagger a+g\sigma_x(a+a^\dagger).
 \label{eq:qrm}
\end{equation}
In the weak-coupling, near-resonant regime, the rotating-wave approximation leads to the Jaynes--Cummings model \cite{JaynesCummings1963}. The counter-rotating terms are part of the QRM at all couplings; the rotating-wave approximation becomes progressively less accurate as the coupling is increased, particularly away from the weak-coupling near-resonant regime. These regimes have been experimentally explored with several quantum platforms.

In neutrino physics, two-state Hamiltonians arise naturally in flavor and spin--flavor evolution. In particular, neutrino magnetic moments generate transitions between opposite-helicity states and, for transition moments, between different mass eigenstates \cite{LimMarciano1988,Pal1988,Akhmedov1988,Studenikin2016}. Separately, periodically varying matter profiles have already been shown to generate Rabi-type neutrino flavor transformation \cite{MaShalgarDuan2018}. The purpose here is different: we construct a quantized-field neutrino Hamiltonian and determine whether its effective operator belongs to the same quantum-Rabi class as the QRM. The subsequent spectral transfer is then performed at the level of the dimensionless Hamiltonian parameters, rather than interpreted as an experimental realization of a neutrino QRM.

There are three neutrino mass eigenstates, so a Dirac description that retains both helicities contains six states. The standard QRM correspondence developed below therefore applies to a selected, approximately isolated two-state spin--flavor channel rather than to the full three-flavor Hamiltonian. The reduction is made explicit below.

This distinction is essential. A formal resemblance between two differential equations is weaker than an operator identity, and an operator-level correspondence is still different from an experimentally demonstrated realization. We therefore proceed in three steps: (i) derive the two-state effective neutrino Hamiltonian, (ii) identify its QRM parameters, and (iii) use an experimentally calibrated QRM as a reference for the corresponding dimensionless parameter space.

\subsection{Three-flavor structure and validity of the two-state reduction}
For three Dirac neutrino mass eigenstates, retaining both helicities gives the basis
\begin{equation}
 \Psi=(\nu_{1L},\nu_{2L},\nu_{3L},\nu_{1R},\nu_{2R},\nu_{3R})^{\mathsf T}.
 \label{eq:full_basis}
\end{equation}
The corresponding effective Hamiltonian has the block form
\begin{equation}
 H_{6\times6}=\begin{pmatrix}
 H_L & H_{\rm mag}\\
 H_{\rm mag}^{\dagger} & H_R
 \end{pmatrix},
 \label{eq:full6}
\end{equation}
where $H_L$ and $H_R$ contain the vacuum mass terms and matter potentials, while $H_{\rm mag}$ contains the transition magnetic moments multiplied by the transverse magnetic field. In the mass basis the vacuum contribution is diagonal, but a flavor-diagonal matter potential is generally not diagonal after transformation to the mass basis. Thus the full three-flavor problem need not decompose into three independent two-state QRM sectors.

For a Dirac neutrino, the right-handed states have negligible standard matter potential in ordinary matter, so $V_{jR}\simeq0$ is appropriate for the selected channel. If only one transition, $\nu_{iL}\leftrightarrow\nu_{jR}$, is relevant and the remaining states are sufficiently far detuned or negligibly coupled over the propagation region, the full Hamiltonian reduces to the invariant two-state sector
\begin{equation}
 \mathcal H_{ij}=\operatorname{span}\{\lvert\nu_{iL}\rangle,\lvert\nu_{jR}\rangle\}.
 \label{eq:sector}
\end{equation}
Within this controlled reduction, the QRM mapping is exact for the projected two-state Hamiltonian. If several transition moments are simultaneously relevant, the sectors share the same neutrino Hilbert space and are coupled; the resulting problem is then a multilevel spin--boson (generalized Rabi) model rather than the standard two-level QRM.

Accordingly, every quantity below carrying indices $i,j$ refers to one selected channel. The numerical benchmark later in the paper makes the choice explicit: $i=3$, $j=1$, so that $\Delta m^2_{ij}=\Delta m^2_{31}$. The resulting numerical values should therefore be read as a benchmark for the $\nu_{3L}\leftrightarrow\nu_{1R}$ channel, not as a calculation of all three neutrino flavors simultaneously.

Consider the spin--flavor sector
\begin{equation}
 \lvert 1\rangle=\lvert\nu_{iL}\rangle,\qquad \lvert 2\rangle=\lvert\nu_{jR}\rangle.
 \label{eq:states}
\end{equation}
For an ultrarelativistic neutrino,
\begin{equation}
 E_i\simeq p+\frac{m_i^2}{2p},\qquad
 E_j\simeq p+\frac{m_j^2}{2p},\qquad (E\simeq p).
 \label{eq:ultra}
\end{equation}
Including effective matter potentials gives
\begin{equation}
 H_0=\begin{pmatrix}
 \dfrac{m_i^2}{2E}+V_{iL}&0\\[5pt]
 0&\dfrac{m_j^2}{2E}+V_{jR}
 \end{pmatrix}.
 \label{eq:h0matrix}
\end{equation}
After subtracting the trace,
\begin{equation}
 H_0=\frac{\Delta_s}{2}\sigma_z,
 \label{eq:h0}
\end{equation}
with
\begin{equation}
 \Delta_s=\frac{m_i^2-m_j^2}{2E}+V_{iL}-V_{jR}.
 \label{eq:deltas}
\end{equation}
For a Dirac right-handed state, $V_{jR}\simeq0$ in ordinary matter.

The transverse magnetic-moment interaction requires one physical qualification before the field is quantized. At the level of the effective two-state sector, the electromagnetic dipole interaction can be written schematically as
\begin{equation}
 {\cal L}_{\rm em}
 =\frac{1}{2}\,\overline{\nu_i}
 \sigma_{\mu\nu}\left(\mu_{ij}-i\epsilon_{ij}\gamma_5\right)\nu_j
 F^{\mu\nu}+\mathrm{h.c.},
 \label{eq:em_lagrangian}
\end{equation}
where the precise relation between the magnetic and electric transition moments depends on the neutrino type and phase conventions. For the present spin--flavor channel we retain the magnetic transition term and project onto the basis in Eq.~\eqref{eq:states}. In the ultrarelativistic limit, and after choosing the relative phase of the two basis states so that the transition matrix element is real, the transverse component gives the effective two-level interaction
\begin{equation}
 H_B=\mu_{ij}B_\perp\sigma_x,
 \label{eq:hbclassical}
\end{equation}
up to an irrelevant overall sign. This is the effective spin--flavor-precession structure used in the standard two-state treatment \cite{LimMarciano1988,Pal1988,Akhmedov1988}. Equation~\eqref{eq:hbclassical} should therefore be understood as an effective projection of the electromagnetic interaction, not as a claim that the full relativistic neutrino field theory contains only two states.

For a single electromagnetic mode,
\begin{equation}
 H_{\rm EM}=\omega_c a^\dagger a,
 \qquad [a,a^\dagger]=1,
 \label{eq:hem}
\end{equation}
and the transverse magnetic field may be represented at the neutrino trajectory as
\begin{equation}
 \hat B_\perp=B_{\rm zpf}(a+a^\dagger),
 \label{eq:bzpf}
\end{equation}
where $B_{\rm zpf}$ is the zero-point magnetic-field amplitude of the selected mode. Hence
\begin{equation}
 H_B=\mu_{ij}B_{\rm zpf}\sigma_x(a+a^\dagger).
 \label{eq:hbquant}
\end{equation}
Defining
\begin{equation}
 g_\nu=\mu_{ij}B_{\rm zpf},
 \label{eq:gnu}
\end{equation}
the complete effective Hamiltonian is
\begin{equation}
 H_\nu=\frac{\Delta_s}{2}\sigma_z+\omega_c a^\dagger a+g_\nu\sigma_x(a+a^\dagger).
 \label{eq:hnu}
\end{equation}

Comparison of Eqs.~\eqref{eq:qrm} and \eqref{eq:hnu} gives the operator dictionary
\begin{equation}
 \omega_0\leftrightarrow\Delta_s,\qquad
 g\leftrightarrow g_\nu=\mu_{ij}B_{\rm zpf},\qquad
 \omega_c\leftrightarrow\omega_c.
 \label{eq:dictionary}
\end{equation}
The two-level basis maps as $\lvert\nu_{iL}\rangle,\lvert\nu_{jR}\rangle\leftrightarrow\lvert g\rangle,\lvert e\rangle$, up to labeling. This is a correspondence between effective two-level Hilbert spaces; it does not identify a neutrino physically with an atomic two-level system.

The more useful statement for the present work follows after division by $\omega_c$:
\begin{equation}
 \frac{H_\nu}{\omega_c}
 =\frac{1}{2}\frac{\Delta_s}{\omega_c}\sigma_z+a^\dagger a
 +\frac{g_\nu}{\omega_c}\sigma_x(a+a^\dagger).
 \label{eq:dimensionless}
\end{equation}
Thus the normalized spectrum depends on the two dimensionless parameters
\begin{equation}
 r_\nu=\frac{\Delta_s}{\omega_c},\qquad
 \eta_\nu=\frac{g_\nu}{\omega_c}.
 \label{eq:dimensionlessparams}
\end{equation}
A genuine spectral correspondence with an experimental QRM spectrum requires matching these dimensionless parameters, not merely matching one oscillation frequency.

Writing $\sigma_x=\sigma_++\sigma_-$ gives
\begin{equation}
 H_{\rm int}=g_\nu(\sigma_+a+\sigma_-a^\dagger+\sigma_+a^\dagger+\sigma_-a).
 \label{eq:fourterms}
\end{equation}
Near resonance, $|\Delta_s-\omega_c|\ll\Delta_s+\omega_c$, and for sufficiently weak coupling, the counter-rotating terms may be neglected. The result is
\begin{equation}
 H_{\rm JC}=\frac{\Delta_s}{2}\sigma_z+\omega_c a^\dagger a+g_\nu(\sigma_+a+\sigma_-a^\dagger).
 \label{eq:jc}
\end{equation}
The photon-number-dependent coupling is proportional to $g_\nu\sqrt{n+1}$. The integer $n$ is a bosonic occupation number and should not be identified with a neutrino oscillation index.

\section{Experimental benchmark and Juddian spectrum}
A particularly relevant experimental benchmark is the trapped-ion quantum simulation of the QRM by Lv \emph{et al.} \cite{Lv2018}. The experiment tuned the simulated coupling and explored the ultrastrong and deep-strong coupling regimes. Importantly for the present purpose, the authors directly measured the energy spectrum of the QRM in the ultrastrong-coupling regime and observed the breakdown of the rotating-wave approximation as the coupling increased.

This makes the experiment a useful experimental reference for the functional dependence of QRM eigenenergies on the dimensionless parameters $\omega_0/\omega_c$ and $g/\omega_c$. We do not interpret the trapped-ion measurement as evidence that a neutrino system has reproduced the same spectrum.

At the QRM resonance condition,
\begin{equation}
 \frac{\omega_0}{\omega_c}=1,
 \label{eq:resonance}
\end{equation}
and the experimental spectra can be parameterized by $g/\omega_c$. In particular, Lv \emph{et al.} report experimental spectra for $g/\omega_m=0.3$ and $0.6$; we use the former as a concrete experimental reference point \cite{Lv2018}. The measured spectrum is obtained from the probe-transition frequencies, while the dimensionless coupling is the experimentally controlled QRM parameter.

To make the transfer explicit, consider the experimentally realized spectral point $g/\omega_c=0.30$ reported by Lv \emph{et al.} \cite{Lv2018}. On the resonant slice $\omega_0/\omega_c=1$, the operator dictionary gives $\Delta_s/\omega_c=1$ and $g_\nu/\omega_c=0.30$. At the quoted benchmark, the trapped-ion parameters are $g/(2\pi)=12.5$ kHz and $\omega_c/(2\pi)=41.7$ kHz. In energy units, $\hbar\omega_c=1.72\times10^{-10}$ eV and $g_\nu=5.17\times10^{-11}$ eV. If this dimensionless point is mapped to a vacuum neutrino sector with $\Delta m^2_{31}=2.513\times10^{-3}$ eV$^2$, the corresponding neutrino energy is $E_\nu=7.3$ MeV. These neutrino quantities are mapped values, not measurements.

\begin{table*}[t]
\caption{Experimental-to-neutrino parameter transfer at the spectral point $g/\omega_c=0.30$ of Lv \emph{et al.}~\cite{Lv2018}. The QRM quantities in the second column are experimentally controlled or measured; the third column is the corresponding neutrino parameter under the exact operator dictionary.}
\label{tab:experimental_mapping}
\centering
\begin{ruledtabular}
\begin{tabular}{lcc}
Quantity & QRM experiment & Neutrino correspondence \\ 
\hline
$g/\omega_c$ & $0.30$ & $g_\nu/\omega_c=0.30$ \\ 
$\omega_0/\omega_c$ & $1$ (resonant) & $\Delta_s/\omega_c=1$ \\ 
$\omega_c/(2\pi)$ & $41.7$ kHz & $41.7$ kHz (mapped) \\ 
$g/(2\pi)$ & $12.5$ kHz & $g_\nu/(2\pi)=12.5$ kHz (mapped) \\ 
$\hbar\omega_c$ & $1.72\times10^{-10}$ eV & $\Delta_s=1.72\times10^{-10}$ eV \\ 
$g$ & $5.17\times10^{-11}$ eV & $g_\nu=5.17\times10^{-11}$ eV \\ 
$E_\nu$ & -- & $7.3$ MeV (mapped) \\ 
$\mu_{ij}$ & -- & $7\times10^{-12}\mu_B$ (illustrative) \\ 
$B_{\rm zpf}$ & -- & $\sim1.3\times10^5$ T$^*$ \\ 
\end{tabular}
\end{ruledtabular}
\begin{flushleft}
\footnotesize $^*$Illustrative field required if $\mu_{ij}\simeq7\times10^{-12}\mu_B$ is used to reproduce the mapped $g_\nu$; this is a feasibility estimate, not an observed field.
\end{flushleft}
\end{table*}

The table should be read as a quantitative parameter-transfer test. The QRM entries are experimentally realized, whereas the neutrino entries are obtained by applying the exact Hamiltonian dictionary. Thus an experimentally measured QRM spectral point defines a definite point in the neutrino parameter space, while the last rows simultaneously expose the physical challenge of realizing the mapped coupling.

\paragraph{Coupling-scale comparison.}
The dimensionless coupling provides a useful way to distinguish the experimentally demonstrated QRM regime from an illustrative neutrino-scale estimate. For the $E_\nu=1$ MeV vacuum benchmark, $|\Delta_s|=1.2565\times10^{-9}$ eV and hence the resonant mode energy is $\hbar\omega_c=1.2565\times10^{-9}$ eV. If the illustrative transition moment $\mu_{ij}=7\times10^{-12}\mu_B$ is combined with a zero-point magnetic field of $1$ T, then
\begin{equation}
 g_\nu=\mu_{ij}B_{\rm zpf}\simeq4.05\times10^{-16}~{\rm eV},
 \qquad
 \eta_\nu=\frac{g_\nu}{\omega_c}\simeq3.2\times10^{-7}.
 \label{eq:one_tesla_scale}
\end{equation}
By contrast, the trapped-ion QRM benchmark used above has $g/\omega_c=0.30$. Reaching that same dimensionless coupling with the illustrative transition moment would require
\begin{equation}
 B_{\rm zpf}\simeq1.28\times10^5~{\rm T}.
 \label{eq:required_bzpf}
\end{equation}
The latter is therefore a required field for the chosen parameter transfer, not an experimentally established electromagnetic field. This comparison does not constitute a bound on $\mu_{ij}$: experimental limits on an effective neutrino magnetic moment do not automatically constrain one specific transition moment. It instead quantifies the separation between an ordinary illustrative field scale and the QRM benchmark used for the cross-platform mapping.

\begin{table*}[t]
\caption{Illustrative coupling-scale comparison for the $E_\nu=1$ MeV vacuum benchmark. The $1$ T row is a derived estimate using the illustrative transition moment; the QRM row is an experimentally realized dimensionless benchmark. The large-field row is the field required to reproduce $g_\nu/\omega_c=0.30$ with the same illustrative moment.}
\label{tab:coupling_scale}
\centering
\begin{ruledtabular}
\begin{tabular}{lcc}
Scenario & $B_{\rm zpf}$ & $g_\nu/\omega_c$ \\ 
\hline
Illustrative neutrino scale & $1$ T & $3.2\times10^{-7}$ \\ 
QRM experimental benchmark & -- & $0.30$ \\ 
Mapped coupling requirement & $1.28\times10^5$ T & $0.30$ \\ 
\end{tabular}
\end{ruledtabular}
\end{table*}

The central point is not that a neutrino experiment has already reproduced the trapped-ion spectrum. Rather, the experimentally realized QRM spectrum supplies a calibrated reference for the same dimensionless spectral problem, and the exact Juddian construction shows how QRM results can be transferred to the neutrino effective Hamiltonian. The operator correspondence is exact within the stated two-state and single-mode approximations; physical realization remains constrained by the transition magnetic moment, the mode geometry, and the coherence of the neutrino--mode interaction.

\begin{equation}
 \omega_c=|\Delta_s|.
 \label{eq:nuresonance}
\end{equation}

The operator correspondence can be connected to the exact solution of the quantum Rabi model. We use the convention employed in the constraint-polynomial literature,
\begin{equation}
 H_{\rm QRM}=\omega_c a^\dagger a+g(a+a^\dagger)\sigma_x+\Delta\sigma_z,
 \label{eq:qrm_braak_convention}
\end{equation}
which differs from Eq.~\eqref{eq:qrm} only by the identification $\Delta=\omega_0/2$.

Braak established that the symmetric QRM is integrable through its conserved parity and provided an exact spectral construction in terms of parity-resolved transcendental functions \cite{Braak2011}. This general solution describes the full spectrum, including generic parameter values for which the eigenvalues are obtained from transcendental spectral conditions rather than finite polynomials. The Juddian states form a special subset of this exact spectrum: at isolated parameter values the corresponding series terminates, yielding finite polynomial solutions and exact level degeneracies between opposite parity sectors. Constraint polynomials provide a particularly convenient algebraic description of these special points \cite{Zhang2012,Kimoto2021}.

In the present work we do not attempt to transfer or analyze the full Braak transcendental solution. Our focus is the Juddian subset because it gives explicit algebraic conditions that can be written directly in terms of neutrino parameters and used to identify concrete spectral loci. Thus Braak's solution supplies the general exact framework, whereas the constraint-polynomial construction supplies the specific part of that framework exploited in this paper. For the symmetric QRM, define
\begin{equation}
 x=\left(\frac{2g}{\omega_c}\right)^2,
 \qquad
 y=\left(\frac{\Delta}{\omega_c}\right)^2.
 \label{eq:xy}
\end{equation}
The order-$N$ Juddian condition is
\begin{equation}
 P_N^{(N)}(x,y)=0,
 \label{eq:judd_condition}
\end{equation}
with
\begin{align}
 P_0^{(N)}&=1,\\
 P_1^{(N)}&=x+y-1,
\end{align}
and, for $k\ge2$,
\begin{equation}
 P_k^{(N)}=(kx+y-k^2)P_{k-1}^{(N)}
-k(k-1)(N-k+1)xP_{k-2}^{(N)}.
 \label{eq:constraint_recurrence}
\end{equation}
At a Juddian point the corresponding exceptional eigenvalue is
\begin{equation}
 E_{J,N}=N\omega_c-\frac{g^2}{\omega_c}.
 \label{eq:judd_energy}
\end{equation}
For the ordinary Hermitian QRM these are parity-sector degeneracies. They should not be confused with non-Hermitian exceptional points, where eigenvalues and eigenvectors coalesce.

The neutrino Hamiltonian in Eq.~\eqref{eq:hnu} is already in the form of Eq.~\eqref{eq:qrm_braak_convention}, with the exact identifications
\begin{equation}
 \Delta\longleftrightarrow\frac{\Delta_s}{2},
 \qquad g\longleftrightarrow g_\nu=\mu_{ij}B_{\rm zpf}.
 \label{eq:braak_dictionary}
\end{equation}
Therefore the QRM Juddian condition transfers directly to the neutrino system:
\begin{equation}
 P_N^{(N)}\left(
 4\left(\frac{\mu_{ij}B_{\rm zpf}}{\omega_c}\right)^2,
 \frac14\left(\frac{\Delta_s}{\omega_c}\right)^2
 \right)=0.
 \label{eq:neutrino_judd}
\end{equation}
Using Eq.~\eqref{eq:deltas}, this may be written entirely in neutrino and environmental parameters as
\begin{equation}
\begin{aligned}
 P_N^{(N)}\Bigg(&4\left(\frac{\mu_{ij}B_{\rm zpf}}{\omega_c}\right)^2,\nonumber\\
 &\frac14\left(\frac{\Delta m^2_{ij}/(2E)+V_{iL}-V_{jR}}{\omega_c}\right)^2\Bigg)=0.
\end{aligned}
\label{eq:fullymapped_judd}
\end{equation}
Equation~\eqref{eq:fullymapped_judd} is the central spectral result of this work: an exact QRM exceptional-spectrum condition has been converted into a condition on a neutrino spin--flavor Hamiltonian.

\paragraph{Lowest constraint polynomials}
The first three symmetric-QRM constraint polynomials generated by Eq.~\eqref{eq:constraint_recurrence} are
\begin{align}
P_1^{(1)}={}&x+y-1,\\
P_2^{(2)}={}&2x^2+3xy-8x+y^2-5y+4,\\
P_3^{(3)}={}&6x^3+11x^2y-54x^2+6xy^2-58xy+108x
\nonumber\\
&+y^3-14y^2+49y-36.
\label{eq:p13}
\end{align}
Higher-order Juddian constraints are generated without ambiguity by the recurrence in Eq.~\eqref{eq:constraint_recurrence}. We retain the first three explicitly because they already display the increasing algebraic structure while the first two are sufficient for the concrete benchmark discussed below.

\paragraph{The first Juddian locus}
For $N=1$, Eq.~\eqref{eq:neutrino_judd} gives
\begin{equation}
4\frac{g_\nu^2}{\omega_c^2}
+\frac{\Delta_s^2}{4\omega_c^2}=1,
\end{equation}
or
\begin{equation}
\Delta_s^2+16g_\nu^2=4\omega_c^2.
\label{eq:n1_locus}
\end{equation}
Thus the first exceptional spectral locus can be solved exactly for the mode frequency:
\begin{equation}
\omega_{c,1}=\frac12\sqrt{\Delta_s^2+16\mu_{ij}^2B_{\rm zpf}^2}.
\label{eq:n1_frequency}
\end{equation}
The associated exceptional energy is obtained from Eq.~\eqref{eq:judd_energy}:
\begin{equation}
E_{\nu,J}^{(1)}=\omega_{c,1}-\frac{g_\nu^2}{\omega_{c,1}}.
\label{eq:n1_energy}
\end{equation}
Using Eq.~\eqref{eq:n1_locus}, this can instead be written as
\begin{equation}
E_{\nu,J}^{(1)}=\frac{12\omega_{c,1}^2+\Delta_s^2}{16\omega_{c,1}}.
\label{eq:n1_energy_simplified}
\end{equation}

\paragraph{The second Juddian locus}
For $N=2$, the transferred constraint is
\begin{equation}
(2q^2+3q+1)y^2-(8q+5)y+4=0,
\label{eq:n2_constraint}
\end{equation}
where it is convenient to define
\begin{equation}
 q=16\frac{g_\nu^2}{\Delta_s^2},
 \qquad
 y=\frac{\Delta_s^2}{4\omega_c^2}.
 \label{eq:q_y}
\end{equation}
Hence
\begin{equation}
 y_{2,\pm}=\frac{8q+5\pm\sqrt{(8q+5)^2-16(2q^2+3q+1)}}
 {2(2q^2+3q+1)},
\label{eq:n2_roots}
\end{equation}
and each positive root determines a Juddian mode frequency through
\begin{equation}
\omega_{c,2,\pm}=\frac{|\Delta_s|}{2\sqrt{y_{2,\pm}}}.
\label{eq:n2_frequency}
\end{equation}
This explicitly demonstrates that the QRM exceptional-spectrum result supplies more than a single resonance condition when transferred to the neutrino system.

\paragraph{Weak-coupling limit of the Juddian loci}
At $g_\nu=0$, the constraint polynomials satisfy
\begin{equation}
P_N^{(N)}(0,y)=\prod_{k=1}^{N}(y-k^2).
\label{eq:weak_poly}
\end{equation}
Consequently, the $N$-th Juddian polynomial has $N$ branches satisfying
\begin{equation}
 y_{N,k}\longrightarrow k^2,
 \qquad k=1,\ldots,N,
\end{equation}
and therefore
\begin{equation}
\omega_{c,N,k}\longrightarrow\frac{|\Delta_s|}{2k}.
\label{eq:weak_frequency}
\end{equation}
This is important because the Juddian spectrum is not confined to ultrastrong coupling. The exceptional loci can approach arbitrarily weak coupling, although increasingly precise detuning is then required.

For example, expanding the positive-$\Delta_s$ branches in $\eta_\nu=g_\nu/\omega_c$ gives
\begin{equation}
\frac{\Delta_s}{\omega_c}=2k+C_{Nk}\eta_\nu^2+O(\eta_\nu^4),
\label{eq:branch_expansion}
\end{equation}
with
\begin{align}
C_{11}&=-4,\qquad
C_{21}=-\frac{20}{3},\qquad C_{22}=-\frac83,\\
C_{31}&=-\frac{28}{3},\qquad
C_{32}=-\frac{56}{15},\qquad C_{33}=-\frac{12}{5}.
\end{align}
These expansions show explicitly how the exact Juddian loci depart from their uncoupled positions. For negative $\Delta_s$, the corresponding branch is obtained by replacing the left-hand side by $|\Delta_s|/\omega_c$. For a negative $\Delta_s$, the corresponding branch is obtained by replacing the left-hand side by $|\Delta_s|/\omega_c$.

For vacuum propagation,
\begin{equation}
 \Delta_s=\frac{\Delta m^2_{31}}{2E_\nu}.
 \label{eq:vacuumdelta}
\end{equation}
Using the atmospheric splitting from the NuFIT global analysis \cite{NuFIT6},
\begin{equation}
 \Delta m^2_{31}=2.513\times10^{-3}\ {\rm eV}^2,
 \label{eq:dm31}
\end{equation}
and the illustrative benchmark $E_\nu=1$ MeV gives
\begin{equation}
 \Delta_s=1.2565\times10^{-9}\ {\rm eV}.
 \label{eq:deltanum}
\end{equation}
The corresponding resonant frequency, $\omega_c=|\Delta_s|$, would be
\begin{equation}
 f_{\rm res}=\frac{|\Delta_s|}{h}\simeq3.04\times10^5\ {\rm Hz}.
 \label{eq:fc}
\end{equation}
This is only a constructed scale; it is not an observed neutrino electromagnetic mode.

For an illustrative zero-point field $B_{\rm zpf}=1$ T and the effective magnetic-moment scale used as a conservative reference in the preceding discussion, the coupling remains extremely small compared with $|\Delta_s|$. Since an effective magnetic-moment limit is not identical to a specific transition moment, we do not interpret it as a direct bound on $g_\nu$. Instead, it establishes the order of magnitude expected if the transition moment is of the same small scale.

In the weak-coupling limit, the corresponding first five branch frequencies are therefore approximately
\begin{equation}
\omega_c\simeq
\frac{|\Delta_s|}{2},\frac{|\Delta_s|}{4},
\frac{|\Delta_s|}{6},\frac{|\Delta_s|}{8},\frac{|\Delta_s|}{10},
\end{equation}
or, in frequency units,
\begin{equation}
f_c\simeq152,\ 76.0,\ 50.6,\ 38.0,\ 30.4~{\rm kHz}.
\label{eq:judd_freqs}
\end{equation}
These values are not measurements. They are the weak-coupling limits of the exact QRM Juddian loci after the parameter map has been applied to the neutrino Hamiltonian.

The associated transferred exceptional energies are
\begin{equation}
E_{\nu,J}^{(N,k)}=N\omega_{c,N,k}
-\frac{\mu_{ij}^2B_{\rm zpf}^2}{\omega_{c,N,k}}.
\label{eq:nu_judd_energy}
\end{equation}
Thus the exact QRM formula for the exceptional energy is also directly applicable to the mapped neutrino Hamiltonian.

The QRM has the conserved parity operator
\begin{equation}
\Pi=\sigma_z(-1)^{a^\dagger a},
\qquad [H_{\rm QRM},\Pi]=0.
\label{eq:parity_exact}
\end{equation}
Because the neutrino Hamiltonian is operator-identical to the QRM under Eq.~\eqref{eq:braak_dictionary}, it possesses the same symmetry,
\begin{equation}
[H_\nu,\Pi]=0.
\label{eq:neutrino_parity}
\end{equation}
The Juddian points are therefore crossings of states belonging to opposite parity sectors. The important transferable result is consequently the whole structure
\begin{equation}
\begin{aligned}
&\text{constraint polynomial}\longrightarrow\text{exceptional eigenvalue}\\
&\hspace{2em}\longrightarrow\text{opposite-parity degeneracy}.
\end{aligned}
\end{equation}
not merely an isolated numerical energy formula.

The trapped-ion experiment of Lv \emph{et al.} demonstrated a controlled realization of the quantum Rabi Hamiltonian and reconstructed its energy spectrum in the ultrastrong and deep-strong coupling regimes \cite{Lv2018}. This experiment therefore provides an experimental reference for the spectral structure of the same operator family. It should not, however, be interpreted as evidence for a neutrino realization of the Juddian spectrum.

The appropriate comparison is dimensionless. Define
\begin{equation}
 r_\nu=\frac{\Delta_s}{\omega_c},
 \qquad
 \eta_\nu=\frac{g_\nu}{\omega_c}.
\end{equation}
The neutrino Hamiltonian becomes
\begin{equation}
\frac{H_\nu}{\omega_c}=a^\dagger a+\frac{r_\nu}{2}\sigma_z
+\eta_\nu\sigma_x(a+a^\dagger),
\end{equation}
while the experimental QRM is described by the same dimensionless operator. Thus a QRM spectral result can be applied to the neutrino problem whenever the corresponding pair $(r,\eta)$ is specified. The Juddian construction above gives the strongest analytic example of this transfer because it converts the exact QRM constraint directly into Eq.~\eqref{eq:fullymapped_judd}.

The illustrative neutrino benchmark considered here lies at $\eta_\nu\ll1$, whereas the trapped-ion experiments access much larger values. The common mathematical operator class therefore spans very different physical parameter regimes; the trapped-ion experiment does not establish the accessibility of the neutrino regime.

\section{QRM benchmarks for quantized neutrino dynamics}
The QRM literature provides experimentally established dynamical and spectral benchmarks that are distinct from the exact Juddian construction. These benchmarks should be interpreted as conditional consequences of the Hamiltonian correspondence, not as established neutrino phenomena. In particular, the trapped-ion realization of Lv \emph{et al.} observed breakdown of the rotating-wave approximation as the simulated coupling was increased, as well as nontrivial entanglement between the bosonic mode and the two-level system \cite{Lv2018}. A later trapped-ion experiment demonstrated parity-symmetry-protected phenomena in an extended QRM and directly connected parity to ground-state entanglement and motional observables \cite{Zhang2025}.

For the neutrino Hamiltonian, the relevant dimensionless coupling is
\begin{equation}
\eta_\nu=\frac{g_\nu}{\omega_c}=\frac{\mu_{ij}B_{\rm zpf}}{\omega_c}.
\label{eq:eta_neutrino_benchmark}
\end{equation}
Writing the interaction as in Eq.~\eqref{eq:fourterms}, the two counter-rotating terms are
\begin{equation}
H_{\rm CR}=g_\nu(\sigma_+a^\dagger+\sigma_-a).
\label{eq:counterrotating}
\end{equation}
Thus a quantized-field neutrino calculation based on the full Hamiltonian predicts corrections to the rotating-wave neutrino dynamics whenever these terms are not negligible. The statement is conditional: it does not imply that ordinary neutrino propagation has been observed to enter a strong-coupling QRM regime.

The same conditional logic applies to entanglement. If the electromagnetic mode is prepared in a nontrivial quantum state and the neutrino remains coherently coupled to it, the state generally takes the form
\begin{equation}
|\Psi(t)\rangle=\sum_n\left[c_{n,L}(t)|\nu_{iL},n\rangle+c_{n,R}(t)|\nu_{jR},n\rangle\right],
\label{eq:nu_field_entanglement}
\end{equation}
so that neutrino spin--flavor and mode occupation are not, in general, separable. The corresponding QRM phenomenon is experimentally established, but its neutrino analogue is a prediction of the quantized-field extension proposed here.

Occupation-dependent dynamics also follow directly from the matrix elements of $a+a^\dagger$. In the rotating-wave limit the transition matrix element between $|\nu_{iL},n\rangle$ and $|\nu_{jR},n+1\rangle$ is $g_\nu\sqrt{n+1}$. Consequently, a superposition of photon-number states produces a superposition of different neutrino transition frequencies. For coherent or otherwise broad field states, this mechanism can generate collapse--revival-type spin--flavor dynamics, as occurs in related QRM and cavity-QED settings. We do not claim such revivals have been observed for neutrinos; they are a calculable consequence that would have to be demonstrated from a specified initial field state, mode geometry, and propagation model.

These observations sharpen the role of the QRM benchmark. The experimentally established QRM results do not supply new measured neutrino parameters. Instead, they identify concrete dynamical signatures that the quantized neutrino Hamiltonian must reproduce if its parameters enter the corresponding QRM regime. The most direct tests are therefore (i) deviations between the full QRM evolution and the rotating-wave approximation, (ii) parity-resolved spectral structure including Juddian crossings, and (iii) correlations between neutrino spin--flavor and electromagnetic-mode occupation.

\section{Physical implications and limitations}
Several qualifications delimit the scope of the correspondence. First, the construction assumes a well-defined single quantized electromagnetic mode along the neutrino trajectory. A realistic calculation must derive $B_{\rm zpf}$ from the mode volume, polarization, boundary conditions, and spatial profile rather than treating it as an arbitrary classical field.

Second, the magnetic moment entering $g_\nu$ should be a transition moment appropriate to the selected spin--flavor channel. An experimental limit on an effective magnetic moment cannot automatically be identified with a limit on a particular $\mu_{ij}$.

Third, the Juddian conditions derived here are spectral conditions of the quantized-field Hamiltonian. They are not claims that a neutrino experiment has observed an exceptional degeneracy. In particular, the term ``exceptional'' is used here in the standard QRM sense of the exceptional/Juddian spectrum and does not imply a non-Hermitian exceptional point.

Finally, the electromagnetic frequencies appearing in Eqs.~\eqref{eq:n1_frequency} and~\eqref{eq:judd_freqs} are mapped spectral scales, not experimentally established neutrino radiation frequencies. A realistic experimental proposal would require an explicit mode and a demonstration that the neutrino remains coherently coupled to that mode over the relevant propagation length.

\section{Conclusions}
We have constructed a two-state effective neutrino spin--flavor Hamiltonian coupled to a single quantized electromagnetic mode and shown that, within the stated two-state and single-mode approximations, it has the operator form of the quantum Rabi model under the explicit parameter correspondence in Eq.~\eqref{eq:braak_dictionary}. The correspondence is therefore an operator-level correspondence within the reduced model, rather than a claim that neutrinos constitute an experimentally realized QRM system.

The exact QRM solution provides the broader mathematical setting for this correspondence. Braak's construction applies to the full spectrum through parity-resolved transcendental equations, whereas the present work deliberately focuses on the Juddian subset, where the exact solution terminates and becomes algebraic. This distinction is important: the Juddian states are not the general QRM spectrum, but they provide especially transparent exact spectral conditions that can be transferred to the neutrino variables. The resulting constraint polynomials, exceptional energies, and opposite-parity degeneracies therefore constitute the main exact results of the present analysis.

In particular, the mapped condition
\begin{equation}
P_N^{(N)}\!\left(
4\frac{\mu_{ij}^2B_{\rm zpf}^2}{\omega_c^2},
\frac{\Delta_s^2}{4\omega_c^2}
\right)=0
\end{equation}
provides a direct algebraic relation among the neutrino splitting, transition magnetic moment, zero-point magnetic field, and mode frequency. The first Juddian locus and the weak-coupling branches give explicit examples of how these exact QRM structures translate into neutrino parameter space.

The QRM benchmarks discussed in this work should be interpreted as conditional consequences of the Hamiltonian correspondence. Counter-rotating corrections, neutrino--field entanglement, and occupation-dependent dynamics follow from the quantized-field model when the corresponding dynamical regime and field state are realized, but they are not presented as already observed neutrino phenomena. Likewise, the trapped-ion QRM experiments provide experimental benchmarks for the operator family, not experimental evidence for a neutrino realization of the mapped Juddian spectrum.

The three-flavor structure defines an additional limitation. The standard QRM mapping applies to an isolated two-state channel within the full six-state Dirac-neutrino system. If several transition moments or matter-induced off-diagonal terms are simultaneously relevant, the resulting problem is generally a multilevel generalized Rabi or spin--boson model rather than a single standard QRM. Consequently, the present results should be understood as exact statements for the selected two-state sector, with the reduction to that sector requiring separate physical justification in any concrete application.

Finally, the mapping alone does not determine an absolute neutrino mass or establish a new magnetic-moment measurement. Its present significance is the identification of an exact spectral correspondence and the transfer of a well-defined class of QRM algebraic structures to neutrino spin--flavor dynamics. A physical realization would require, in addition, a specified quantized electromagnetic mode, an appropriate transition moment, sufficient coherence, and a demonstrable regime in which the reduced two-state description remains valid.

\begin{acknowledgments}
The author acknowledges the use of AI-assisted tools for drafting and language editing. All equations, physical arguments, numerical estimates, and references were reviewed by the author.
\end{acknowledgments}

\end{document}